\documentstyle[11pt]{article}
\titlepage
\newcommand{\p}{\underline}
\newcommand{\fr}{\frac}

\newcommand{\de}{\delta}

\newcommand{\be}{\begin{equation}}

\newcommand{\ee}{\end{equation}}

\newcommand{\cH}{{\cal H}}
\newcommand{\bea}{\begin{eqnarray}}
\newcommand{\eea}{\end{eqnarray}}

\begin{document}


\title
{Dirac and Friedmann Observables \\
in Quantum Universe with Radiation.}

\author{S.A.Gogilidze\thanks{
Permanent address: IHEP, Tbilisi State University,
380086, Tbilisi, Georgia.} , A.M.Khvedelidze\thanks{
Permanent address: Tbilisi Mathematical Institute,
380093, Tbilisi, Georgia.} , V.V. Papoyan\thanks{
Permanent address: Yerevan State University, 375049, Yerevan, Armenia.} ,\\
Yu.G.Palii,
V.N.Pervushin\\[0.3cm]
{\normalsize\it Joint Institute for Nuclear Research},\\
 {\normalsize\it 141980, Dubna, Russia.}}

\date{\empty}

\maketitle
\medskip


\begin{abstract}
{\large

{ Relations between the Friedmann observables of the expanding Universe
and the Dirac observables in the generalized Hamiltonian approach are
established for the Friedmann cosmological model of the Universe with
the field excitations imitating radiation.

A full separation of the physical sector from the gauge
one is fulfilled by the method of the gaugeless reduction in which the
gravitational part of the energy constraint is considered as a new momentum.
We show that this reduction removes an infinite factor from the
Hartle -- Hawking functional integral, provides the normalizability of the
Wheeler -- DeWitt wave function, clarifies its relation to the observational
cosmology, and picks out a conformal frame of Narlikar. }}

\end{abstract}

\vspace{2 cm}

\newpage

\section {\p { Statement of the problem}}

There is a hope  of solving fundamental problems of cosmology of the early
Universe by help of quantum gravity ~\cite{Dir58,ADM,Wheel,DeWitt,Faddeev}.
The problem of quantization has stimulated the development
of the Hamiltonian approach to the theory of gravity and cosmological models
of the Universe.
A lot of papers  and some monographs (see e.g. \cite{Ryan1,Ryan2}) have
been devoted to the Hamiltonian description of cosmological models
of the Universe. The main peculiarity of the Hamiltonian
theory of gravity is the presence of nonphysical variables and constraints.
They arise due to the  diffeomorphism invariance of the theory
which is the basis of the difficulties
with the solution for the important conceptual  problems\\
-  treatment of the observable time in classical cosmology\,\
-  interpretation of the wave function  and its non-normalizability ,\\
-  relations between the observational cosmology (the Hubble law and red
   shift) and the Dirac observables in  the Hamiltonian description of the
    classical and  quantum cosmologies.

One of the possible solution of these problems  in the
Hamiltonian approach is to reduce the initial constraint system to
unconstrained one by separation of pure gauge degrees of freedom from
physical ones.
In the present paper, we would like to apply
recently developed method of the Hamiltonian reduction of singular systems
 with the full separation of the gauge sector ~\cite{PhRevD,JMPh}.
to a simple, but important, cosmological model of the Universe with
scalar field to investigated the problems listed above and to compare
our reduced quantization with the extended approach
~\cite{Wheel,DeWitt,Faddeev}.

The content of the paper is the following. Section 2 is devoted to
observational cosmology.
In Section 3, we present the Lagrangian model the equations of which
coincide with the ones of the Friedmann Universe filled in by radiation.
In Section 4, the gaugeless version of the Dirac Hamiltonian description
~\cite{PhRevD,JMPh} of the model is expounded, and the phase
space reduction is fulfilled by separating
of the physical and nonphysical sectors. In Section 5, we establish the
relation
between the Dirac observables in the Hamiltonian approach and the Friedmann
 ones in the classical cosmology.
Section 6 is devoted to the quantization of the model in the
reduced phase space and the description of the cosmological observables
in quantum theory. In Section 7, the functional integral is constructed
which is adequate to the gaugeless quantization. In Section 8,
we show how to modify the Wheeler -- DeWitt wave function so that it
describes the Friedmann cosmological observables. The conclusion is devoted
to the discussion and physical interpretation of the results.

\section{\p {Observational cosmology.}}

\subsection{Experimental data.}

One of the main facts of the observational cosmology is correlation
between the distance of an astronomical object $(R_F)$ to the Earth
and the red shift $z$ (in $\hbar=c=1$ units)
\be \label{z}
z=\frac{\lambda(T_F)}{\lambda(T_F-R_F)}-1=
\frac{1}{\lambda(T_F)}\frac{d\lambda(T_F)}{dT_F}R_F+...\;\;;  \quad
R_F\ll T_F
\ee
where $\lambda(T_F)$ is wave length of photon radiated by an atom on
the Earth and $\lambda(T_F-R_F)$ wave length of photon radiated by an atom
on an astronomical object at the time $(T_F-R_F)$.

The quantity $
\frac{1}{\lambda(T_F)}\frac{d\lambda(T_F)}{dT_F}=H_o$ is known as
the ``Hubble constant''. The present value of this constant
~\cite{natur1},~\cite{natur2}
$$
H_o\propto (70\pm 15)\;\;\frac{\mbox{km}}{\mbox{s Mpc}},
$$
gives the scale of the observational cosmology.

\subsection{Theoretical interpretation.}  

There are two interpretations of this experimental fact.
The recent theoretical cosmology is based on the Friedmann solution
of equations of general relativity for the case of homogeneous and isotropic
distribution of matter in the Universe ~\cite{Fried}.
It is important to emphasize that classical cosmology uses the
comoving  frame of reference with
the Friedmann -- Robertson -- Walker metric
\be
\label{dsf}  
(ds_F)^2 = dT_F^2-a^2(T_F)\gamma_{ij}dx^idx^j ,
\ee
where $a(T_F)$ is cosmic scale factor,
 $\gamma_{ij}dx^idx^j$ is the metric
of the three-dimensional space of the constant curvature
\be
{}^{(3)}R(\gamma_{ij})=\frac{-6k}{r_o^2} \;;\;\;\;k=0,\pm 1,
\ee
($r_o$  is a parameter characterizing a ``size''
of the Universe).
As the consequence of such a choice, one supposes that, in cosmology,
physically measured quantities are the ones which evolve in the proper
(Friedmann) time $T_F$.
The measured quantity of the metric (\ref{dsf}) is the distance
$R_F(T_F)$ to cosmic objects:
\be \label{dist}
R_F(T_F)=a(T_F)R_c\;\;\;\;
R_c=\int\limits\frac{dr'}{\sqrt{1-kr'^2/{r_o^2}}}.
\ee
and the ``Hubble constant''
\be \label{H_o}
H_o=\frac{1}{\lambda(T_F)}\frac{d\lambda(T_F)}{dT_F}R_F
=\frac{1}{a(T_F)}\frac{da(T_F)}{dT_F}.
\ee

The alternative treatment of Hubble law was developed by
Narlikar (see review \cite{Narlik} and the literature cited therein).
According to Narlikar the measured quantity
is the distance  $R_c$
\be\label{rc}
R_c=\int\limits_{o}^{T_c(T_F)}\frac{dT_F}{a(T_c)}=T_c(T_F)
\ee
in the conformal metric
\be \label{dsc}
(ds_c)^2=dT_c^2-\gamma_{ij}dx^idx^j.
\ee

In the conformal (Narlikar) frame
the Universe is stationary and
the ``conformal wave length'' of a photon does not change
during the time of the photon flight
from a "star" to the Earth.
However, the `` conformal mass''
\be\label{mc}
m_c(T_c)=m_Fa(T_c)
\ee
is time dependent and this leads to the  red shift.
In result the Hubble law has the form

$$
z=\fr{m_c(T_c)}{m_{c}(T_c-R_c)}-1.
$$


\section{\p {Model.}}

We begin from the Einstein -- Hilbert action  with the
conformal scalar field
\be\label{WPhi}
W=\int\limits_{}^{}d^4x\sqrt{-g}\left[-\frac{{}^{(4)}R(g_{\mu\nu})}{16\pi G}
\left(1-\frac{16\pi G}{12}\Phi^2\right)+\frac{1}{2}
g^{\mu\nu}\partial_\mu\Phi\partial_\nu\Phi\right].
\ee
The Hamiltonian formulation of gravity is fulfilled in the ADM metric
 ~\cite{ADM}
\be \label{adm}
(ds_E)^2=N^2dt^2-g_{ij}\check dx^i\check dx^j\;;\;\;\;\;\;
\check dx^i=dx^i+N^idt.
\ee

In order to derive a set of equations which is completely equivalent
to the Friedmann -- Einstein ones we choose the metric
\be   
(ds)^2=a^2(t)[N_c^2dt^2-\gamma_{ij}dx^idx^j].
\ee
and the ansatz for the scalar field
\be
\Phi=\fr{\varphi(t)}{a(t)}.
\ee
Instead of eq. (\ref{WPhi}) we get the action in the homogeneous
approximation
\bea
W^F&=&\int\limits_{t_1}^{t_2}dt\left[-\beta\left(\frac{\dot a^2}{2N_c}-
\frac{ka^2}{2r_o^2}N_c\right)+V_{(3)}\left(\frac{\dot \varphi^2}{2N_c}-
\frac{k\varphi^2}{2r_o^2}N_c\right)+\right.\nonumber\\
&& + \left.\frac{\beta}{2}\frac{d}{dt}\left(\frac{\dot a a}{N_c}\right)\right].
\label{Wfi}
\eea
We retained here one of the total derivatives arising from the
the gravitational part of the action (\ref{WPhi}));
$V_{(3)}$ is the volume of
the three-dimensional space with the constant curvature and $\beta$ is a
constant coefficient
\be
\beta=V_{(3)}\frac{6}{2\pi G}\;;\;\;\;\;V_{(3)}|_{k=+1}=2\pi^2r_o^3.
\ee
One can easily be convinced that the set of equations of the model (\ref{Wfi})
is equivalent to the Friedmann Universe filled in by the matter with
the equation of state of the radiation ~\cite{Staniuk}.

The variation of the action (\ref{Wfi}) with respect to the matter field
leads to equation of motion
$\varphi$
\be\label{wfi}
\frac{\delta W}{\delta\varphi}=0\;\;\;\Rightarrow\;\;\;\;\;
-\frac{d}{N_cdt}\left(\frac{d\varphi}{N_cdt}\right)-\frac{k\varphi}{r_o^2}=0.
\ee
The consequence of this equation is the integral of motion
\be\label{efi}
E_c(\varphi)=V_{(3)}\left[\fr{1}{2}\left(\frac{d\varphi}{N_cdt}\right)^2 +
\frac{k\varphi^2}{2r_o^2}\right]\;;\;\;\;\;\frac{d}{dt}E_c(\varphi)=0,
\ee
which plays a role of the conformal energy $E_c$ for the massless
scalar field.

The equation on the variable $N_c$ coincides with the known Einstein
balance of energy of expanding space and matter
\be\label{Nc}
\frac{\delta W}{\delta N_c}=0\;\;\;\Rightarrow\;\;\;\;\;
\beta\left[\left(\frac{da}{N_cdt}\right)^2 +
\frac{ka^2}{2r_o^2}\right]=E_c(\varphi).
\ee
The Friedmann evolution results from equations (\ref{wfi}), (\ref{efi}) and
(\ref{Nc}) when the convention about the
definition of the proper time of observing (5)
\be\label{TF}
dT_F=aN_cdt=adT_c
\ee
is added to these equations.
Substituting eq.(\ref{TF}) into (\ref{Nc}) and solving this equation under
$T_F$, we get the Hubble law of the radiation dominant Universe in the
parametric form
\be \label{aTc}
a(T_c)=\sqrt{\frac{2E_c(\varphi )r_o^2}{\beta}}S_k\left(\frac{T_c}{r}\right)
\;;\;\;\;\;T_F(T_c)=\int\limits_{o}^{T_c}dT_ca(T_c)
\ee
where
\be
S_{k=1}(\eta)=\sin \eta\;;\;\;
S_{k=-1}(\eta)=\sinh \eta\;;\;\;
S_{k=0}(\eta)=\eta
\ee
Our problem is to find out the connection between the cosmological
observables and the Dirac observables of the
Hamiltonian approach to the model (\ref{Wfi}) and establish a bridge
between the classical evolution and the wave function of the Universe
determined by the WDW equation
~\cite{Wheel,DeWitt}
\be\label{WDW}
[-\fr{1}{2a\beta}\fr{d^2}{da^2}+\beta\fr{ka}{2r_o^2}-\fr{E_c(\varphi)}{a}]
\Psi_{WDW}(a,\varphi)=0
\ee
which is the quantum analogy of the energy balance equation (17).


\section{\p {Gaugeless Hamiltonian reduction.}}

According to the Dirac classification ~\cite{Dirac} the action (\ref{Wfi})
is a singular. Following to the generalized Hamiltonian approach to
singular theories  this action can be rewritten in the form
\be\label{wham}
W^F[p_\varphi,\varphi;p_a,a]=\int\limits_{}^{}dt
\left\{p_\varphi\dot\varphi-\left[p_a\dot a-\frac{1}{2}\frac{d}{dt}
(p_aa)\right]-N_c\cH_{Ec}\right\},
\ee
where
\be\label{hec}
\cH_{Ec}=-\left(\frac{p_a^2}{2\beta}+\frac{ka^2}{2r_o^2}\beta\right)+\cH_\varphi
\ee
is the conformal version of the Einstein energy and
$$
\cH_{\varphi}=\left(\frac{p_\varphi^2}{2V_{(3)}}+
\frac{k\varphi^2}{2r_o^2}V_{(3)}\right)
$$
is the part describing a homogeneous scalar field (matter).

The considered model (\ref{wham}) faces principal difficulties of
the theory of gravity. The main of these difficulties is the presence
of nonphysical (ignored) variables.
In the phase space $p_\varphi,\varphi;p_a,a,$
one of the momenta depends on the others due to the constraint
$$
      \cH_{Ec}=0.
$$

Let us  discuss the Hamiltonian reduction in the case when
an independent variable is chosen as a matter momentum.
For the complete separation of the physical sector from the nonphysical one,
we apply the method developed in papers ~\cite{PhRevD,JMPh}.
In accordance with this method, such a separation can be fulfilled
using the canonical transformation to new
variables
\be\label{puas}
(p_a,a)\;\;\ss\;\;(\Pi_a,\eta_a),
\ee
so that the gravitation part of the constraint for these variables becomes
a new momentum
\be\label{pi}
\frac{p_a^2}{2\beta}+\frac{ka^2}{2r_o^2}\beta=\Pi_a.
\ee

There are two possible canonical transformations
\be\label{pa}
p_{a(\pm)}=\pm\sqrt{2\beta\Pi_a}C_k(\eta_a)\;;\;
a_{(\pm)}=\pm\sqrt{\frac{2\Pi_a r_o^2}{\beta}}S_k(\eta_a)
\ee
where
\be
(C_{+1}(\eta_a)=\cos\eta_a\;;\;
C_{-1}(\eta_a)=\cosh\eta_a\;;\;
C_{0}(\eta_a)=1).
\ee
It is interesting to note,
that the surface term of the gravitational part of the Einstein -- Hilbert
action (\ref{wfi}) is completely absorbed by the new canonical structure
~\cite{PhLet}
\be
-(p_a\dot a-\frac{d}{dt}(p_aa))=\mp\Pi_a\dot\eta_a r_o.
\ee
In terms of the new variables (\ref{pa}) the action (\ref{wham})
reads
\be\label{WP}
W^F_{(\pm)}(\Pi_a,\eta_a,p_\varphi,\varphi,N_c)=
\int\limits_{}^{}dt\left[p_\varphi\dot\varphi\mp\Pi_a\dot\eta_a r_o-
N_c(-\Pi_a+\cH_\varphi)\right]
\ee
Expression (\ref{WP}) leads to the Hamiltonian equation describing the
nonphysical sector of the variables $(\Pi_a,\eta_a)$
\be
\frac{\delta W^F_{(\pm)}}{\delta \eta_a}=0\;\;\;\Rightarrow\;\;\;\;\;
\pm\dot\Pi_a=0
\ee

\be\label{np2}
\frac{\delta W^F_{(\pm)}}{\delta \Pi_a}=0\;\;\;\Rightarrow\;\;\;\;\;
r_od\eta_a=\pm N_cdt
\ee
and the physical (by the Dirac definition \cite{Dirac}) one
\be
\frac{\delta W^F_{(\pm)}}{\delta p_\varphi}=0\;\;\;\Rightarrow\;\;\;\;\;
\frac{d\varphi}{N_cdt}=\pm\{\cH_\varphi,\varphi\}.
\ee

\be
\frac{\delta W^F_{(\pm)}}{\delta \varphi}=0\;\;\;\Rightarrow\;\;\;\;\;
\frac{dp_\varphi}{N_cdt}=\pm\{\cH_\varphi,p_\varphi\}
\ee
From equation (\ref{np2}) we can see that after transformation (\ref{puas})
the new ignored variable $\eta_a$
turns into the  parameter of time of the evolution of
the Dirac physical variables in the reduced phase space $p_\varphi,\varphi$.
This parameter is invariant under the time-reparametrization group
transformations of the initial time $t$ which is not observable.
We can call the parameter $\eta_a$ {\it the Dirac observable time}.
The role of the Dirac Hamiltonian in the reduced space is played
by the matter part of the Einstein Hamiltonian ${E_c}$ (\ref{hec}) which
coincides with the conventional definition of the matter Hamiltonian
in the flat space.
For the description of the Dirac physical sector, we can restrict ourselves
to the action obtained from (\ref{WP}) by the substitution of the
constraint
\be \label{constr}
\Pi_a=\cH_\varphi  \;.
\ee
As a result, we get the reduced action
\be\label{wred}
W^F_{\pm}|_{\Pi_a=\cH_\varphi}=W^{Red}_\pm=
\int\limits_{t_1}^{t_2}\left(p_\varphi d\varphi\mp\cH_\varphi r_od\eta_a\right)
\ee
which describes excitations of the scalar field in a cavity of the conformal
space with the constant metric $\gamma_{ij}$.

Thus, instead of the extended phase space $N, P_N; a, p_{a; \varphi},
p_{\varphi}$ and the initial action invariant under reparametrizations of the
coordinate time $(t \longmapsto t'=t'(t))$, we have got the reduced phase
space which contains only the fields of matter and the reduced action with
the conformal Hamiltonian $\cH_\varphi$ describing the evolution
of these fields in the stationary conformal space (\ref{dsc})
with respect to the conformal time. All these quantities are
invariant under the coordinate time reparametrizations and can be called
the Dirac observables \cite{Dirac}, including the conformal time.

Our main conclusion is the following:
the gaugeless Hamiltonian reduction, satisfying the correspondence principle,
leads to the Narlikar conformal frame of reference ~\cite{Narlik}
where the observable space seems  stationary and the observable time
$(T_c=r_o\eta_a)$ is monotonic for all types of the space
~\cite{Towmas,Khved}.


\section{\p {Construction of the Friedmann  }
\newline
\p { observables in the Hamiltonian scheme.}}

The Friedmann evolution of the Universe is based on the Einstein convention
about the observable (proper) time (\ref{dsf})
\be \label{ft}
dT_F=a(t)N_cdt\equiv a(\eta_a)r_od\eta_a,
\ee
proper distance
\be \label{fd}
R_F=a(\eta_a)R_c,
\ee
and proper energy
\be\label{en}
E_F=\frac{E_c}{a(\eta_a)}.
\ee
Such an evolution is described by the quantity $a(\eta_a)$
defined through the  canonical transformation
(\ref{pa}) on the constraint surface
(\ref{constr})
\be \label{apm}
a_{(\pm)}=\pm \sqrt{\frac{2E_c r_o^2}{\beta}}S_k(\eta_a),
\ee
where $E_c$ is a value of the energy.


\section{\p {Quantization in the reduced phase space.}}

As in the case of a relativistic particle, two solutions of the energy
constraint corresponding to two reduced actions $W^{Red}_{+}$, $W^{Red}_{-}$
mean that the total
wave function of the Universe represents the superposition of two
wave functions constructed from these actions
\be
\Psi_{Red}(\eta_a,\varphi)=A^+\Psi^{(+)}_{Red}(\eta_a,\varphi)+
A^-\Psi^{(-)}_{Red}(\eta_a,\varphi).
\ee
The functions $\Psi^{(\pm)}$ satisfy the Schr\" odinger equations
\be
\pm\frac{1}{ir_od\eta_a} \Psi^{(\pm)}_{Red}=\hat\cH_\varphi
\Psi^{(\pm)}_{Red}(\eta_a,\varphi),
\ee
and the coefficients $A^+,A^-$ can be treated as
creation  operators of the Universe and anti-Universe ~\cite{Rubakov}.

The wave function $\Psi^{(\pm)}(\eta_a,\varphi)$ can be represented in the form
of the spectral representation over the complete set of eigenfunctions
$<\varphi|n>$ of the reduced Hamiltonian
\bea \label{spectr1}
\Psi^{(+)}_{Red}(\eta_a,\varphi)&=&\sum_{\varepsilon (n)}^{}
e^{-i\varepsilon(n)\eta_a r_o}<\varphi|n>
\eea

\bea \label{spectr2}
\Psi^{(-)}_{Red}(\eta_a,\varphi)&=&\sum_{\varepsilon (n)}^{}
e^{+i\varepsilon(n)\eta_a r_o}<\varphi|n>
\eea
where $<\varphi|n>$ satisfies the equations
\bea
&&\hat\cH_\varphi<\varphi|\eta_a>=\varepsilon(n)<\varphi|\eta_a>,\\
&&\int\limits_{}^{}d\varphi <n_1|\varphi><\varphi|n_2>^*=\delta_{n_1n_2},
\eea
$n$ being set of conserved quantum numbers.
There are two Universes the evolution of which is governed by the invariant
parameter of the conformal time $\eta_a$.
In the following, we shall consider the case of a closed space
$k=+1$, where $<\varphi|n>$ are the Hermite polynomials, and
\be \label{psid}
\varepsilon(n)=\frac{1}{r_o}(n+\frac{1}{2})\;;\;\;\;\;n=0,1,2,3,\ldots \;.
\ee

The spectral decompositions (\ref{spectr1}), (\ref{spectr2}) represent
wave functions of quantum excitations of the massless scalar field,
in a closed cavity of the conformal space. The wave lengths of the excitations
(~and the region of validity of quantum theory~) coincide with the size
of the space occupied by the Universe.

Let us show that the obtained wave function (\ref{spectr1}), in contrast with
the WDW one (\ref{WDW}), has a direct relation to the Friedmann classical
evolution with respect to the Friedmann observable time (\ref{ft})
\be \label{ft2}
dT_F^{\pm}=a_\pm(\eta_a)r_od\eta_a,
\ee
where the scale factor $a$ is expressed through the parameter
$\eta_a$ by the formula of the canonical transformation (\ref{pa})
\be \label{ared}
a_{(\pm)}(\eta_a)=\pm\sqrt{\frac{2\varepsilon(n) r_o^2}{\beta}}\sin(\eta_a),
\ee
for each term of the spectral decomposition (\ref{spectr1}).
Taking into account the connection of the Friedmann time (\ref{ft2}) with
the conformal one $T_c=\eta_ar_o$, we can verify that the result of the
variation of the wave function (\ref{spectr1}) with respect to the
Friedmann time determines the Friedmann observable energy of the red shift
$E_F=\pm\frac{\varepsilon(n)}{a_\pm}$ (7)
for each term of the spectral decomposition (\ref{spectr1})
\bea
-\frac{d}{idT_F^\pm(a)}\Psi^{(\pm)}_{Red}(\eta_a,\varphi)&=&
{\left(\frac{dT_F(a)}{d\eta_a(a)}\right)}^{-1}\frac{d}{id\eta_a}
\Psi^{(\pm)}_{Red}(\eta_a,\varphi)=\nonumber\\
&&=\sum_{\varepsilon (n)}^{} \frac{\varepsilon(n)}{a_\pm}
e^{-i\varepsilon(n)\eta_a r_o}<\varphi|n>.\label{psi}
\eea
We can see that the wave functions (\ref{spectr1}), (\ref{spectr2})
constructed by the gaugeless reduction have the correct correspondences
with the classical evolutions, in both the frames of reference:
the Einstein (\ref{dsf}) and Narlikar (\ref{dsc}).
In the conformal frame (\ref{dsc}), the Dirac observables coincide with
the cosmological ones. In the Einstein frame (\ref{dsf}), the cosmological
observables connected with the Dirac ones by the conformal transformations
with the cosmic scale factor $a$.


\section{\p {Functional integral} }

To connect the reduced wave function with the WDW one (\ref{WDW})
it is useful to write the spectal decomposition of the Green function
\bea
G(\eta_1,\eta_2|\varphi_1,\varphi_2)&=&\sum_{\varepsilon(n)}^{}
\left[e^{i\varepsilon(n)(\eta_1-\eta_2)}<\varphi_1|n><n|\varphi_2>^*+
\right.\nonumber  \\
&&\left.+
e^{-i\varepsilon(n)(\eta_1-\eta_2)}<\varphi_1|n>^*<n|\varphi_2>\right]
\eea
in the form of the functional integral over the variables of the Dirac
physical sector:
\be
G(\eta_1,\eta_2|\varphi_1,\varphi_2)=
\int\limits_{\varphi_{\eta_1}=\varphi_1}^{\varphi_{\eta_2}=\varphi_2}
D\varphi Dp_\varphi \left[
e^{+iW_+^{Red}(p_\varphi,\varphi)}+
e^{iW_-^{Red}(p_\varphi,\varphi)}\right],
\ee
where $W^{Red}_\pm$ are given by formula (\ref{wred})
\be\label{wr}
W^{Red}_\pm=\int\limits_{\eta_1}^{\eta_2}d\eta_a\left(
p_\varphi\frac{d\varphi}{d\eta_a}\mp\cH_\varphi r_o\right),
\ee
and the role of time is played by the parameter $\eta_a$.
Formally, we can return to the initial time coordinate $t$ so that
$\varphi_1=\varphi(t_1)$, $\varphi_2=\varphi(t_2)$ and the action (\ref{wr})
has the form
\be
W^{Red}_\pm=\int\limits_{\eta_a(t_1)=\eta_1}^{\eta_a(t_2)=\eta_2}dt\left(
p_\varphi\frac{d\varphi}{dt}\mp\cH_\varphi r_o\frac{d\eta_a}{dt}\right)
\ee
In the following, we should take into account that $\eta_a$ is not the
variable but the parameter.
With the help of the functional $\de$-function
\be
\int\limits_{}^{}D\Pi_a\delta(-\Pi_a+\cH_\varphi)=\int\limits_{}^{}
DN_cD\Pi_a\exp\left(i\int\limits_{t_1}^{t_2}dtN_c(-\Pi_a+\cH_\varphi)\right)
\ee
one can also introduce additional integration and restore
 all variables of the extended phase space except for
the ignored "variable"  $\eta_a$
\be\label{green}
G(\eta_1,\eta_2|\varphi_1,\varphi_2)=
\int\limits_{\eta_1,\varphi_1}^{\eta_2,\varphi_2}
D\varphi Dp_\varphi D\Pi_a DN_c\left[
e^{+iW_+^F}+e^{iW_-^F}\right],
\ee
where the actions $W^F_\pm$ are defined by eq. (\ref{WP}).
Integration over the ignored "variable"  $\eta_a$ results in
the infinite gauge factor
\be\label{D}
\int\limits_{t_1<t<t_2}^{}D\eta_a(t)=\triangle.
\ee
This factor can be removed by introducing a $\de$-function
\be\label{de}
\int\limits_{t_1<t<t_2}^{}D\eta_a(t)\delta(\eta_a(t)).
\ee
We call this additional constraint the "canonical" gauge. Let us insert
(\ref{de}) into (\ref{green}) and make the transformation to initial
variables $(p_a,a)$
\be
\Pi_a(p_a,a)=\left(\frac{p_a^2}{2\beta}+\frac{ka^2}{2r_o^2}\beta\right)\;;\;\;
\left.\eta_a(p_a,a)\right|_{k=1}=\arctan\left(\frac{\beta a}{r_o^2p_a}\right).
\ee
Two items in (\ref{green}) can be joined keeping in mind that
$$
\begin{array}{l}
\delta\left(\frac{p_a^2}{2\beta}-(\Pi_a-\frac{a^2}{2r_o^2}\beta)\right)=\\
\sqrt{2\beta}\left[\delta\left(p_a-\sqrt{\Pi_a-\frac{a^2}{2r_o^2}\beta}\right)+
\delta\left(p_a+\sqrt{\Pi_a-\frac{a^2}{2r_o^2}\beta}\right)\right].
\end{array}
$$
As a result, we got a functional integral in the Faddeev -- Popov form
~\cite{Faddeev} in the canonical gauge (\ref{de})
  \bea
&&G(\eta_1,\eta_2|\varphi_1,\varphi_2)=\nonumber\\
&&\int\limits_{\eta_1,\varphi_1}^{\eta_2,\varphi_2}
D\varphi Dp_\varphi Dp_a Da DN_c\delta
\left[\arctan\left(\frac{\beta a}{r_o^2p_a}\right)\right]
e^{iW^F(p_\varphi,\varphi;p_a,a;N_c)}.\label{fint}
\eea
We remind that the naive functional integral over the whole phase space
has the form
\be
G(a_1,a_2|\varphi_1,\varphi_2)=
\int\limits_{\varphi_1,a_1}^{\varphi_2,a_2}
Dp_\varphi D\varphi DN_c Dp_a Da e^{iW^F_{ADM}}
\ee
and differs from the canonical result (\ref{fint}) as follows:

i) by the infinite gauge factor (\ref{D}),

ii) by the surface term
\be \label{surf} 
W^F_{ADM}=W^F-\int\limits_{}^{}dt\left[\frac{1}{2}
\frac{d}{dt}(p_aa)\right].
\ee

Just the same naive functional integral including the infinite gauge factor
has been discussed by Hartle and Hawking ~\cite{hart} and until now is used
by many others (see, for example, ~\cite{Hyward}).


\section{\p {Wheeler -- DeWitt equation}}

In this section we shall discuss the connection
between the obtained wave function  (\ref{spectr1}), (\ref{spectr2})
and the solution for the Whee\-ler -- DeWitt equation (\ref{WDW}).

A conformal version of this equation has the following form:
\be  \label{weq}
\left[-\left(\frac{p_a^2}{2\beta}+\frac{ka^2\beta}{2r_o^2}\right)+
\hat\cH_\varphi\right]\Psi_{WDW}(\varphi,a)=0.
\ee
The conserved quantum number $(\varepsilon_n)$
corresponds to  classical integral of motion $(\cH_\varphi)$.
Then, factorization of the wave function takes place

\be
\Psi_{WDW}(\varphi,a)=\sum_{\varepsilon(n)}^{}
\Psi(a,\varepsilon(n))<n|\varphi>,
\ee
and $\Psi(a,\varepsilon_n)$ satisfies the equation
\be\label{udw}
\left[-\left(\frac{p_a^2}{2\beta}+\frac{ka^2\beta}{2r_o^2}\right)+
\varepsilon(n)\right]\Psi(a,\varepsilon(n))=0.
\ee
A solution of this equations will coincide with the wave functions
of the Hamiltonian reduction (\ref{spectr1}), (\ref{spectr2})
if we use the ordering of the operators in $(\ref{udw})$, so that
the momentum $(\hat p)$ acts later than variable $(a)$,
and add a phase multiplier resulting from the surface term (\ref{surf})
contained in the initial Einstein -- Hilbert action.

The prescribed ordering rule leads to the solution
\be
\Psi_{WDW}=A^+\Psi^+_{WDW}+A^-\Psi^-_{WDW}
\ee
\be
\Psi^\pm_{WDW}(a,\varepsilon(n))=\exp\left[
\pm i\sqrt{2\beta}\int\limits_{o}^{a}da'
\sqrt{\varepsilon(n)-\frac{a'^2\beta}{2r_o^2}}\right]
\ee
\be
\hat p_a \Psi^\pm_{WDW}=\pm\sqrt{2\beta}
\sqrt{\varepsilon(n)-\frac{a^2\beta}{2r_o^2}}\Psi^\pm_{WDW}.
\ee
Taking into account the phase multiplier with the phase
formed by the surface term in eq. (\ref{surf})
\be
S(a)=\frac{\sqrt{2\beta}}{2}a
\sqrt{\varepsilon(n)-\frac{a^2\beta}{2r_o^2}}
\ee
leads to the relation between the wave functions: the reduced one
(\ref{spectr1}) and the WDW function (\ref{WDW})
\be
\Psi^\pm_{Red}(a,\varphi)=e^{\pm i\varepsilon(n)r_o\eta_a
(a)}<\varphi|n>=
e^{\mp iS(a)}\Psi^\pm_{WDW}(a,\varepsilon(n))<\varphi|n>.
\ee
As the variable $a$ turns into a parameter, we can demand
normalizability of this function only for the variables of the reduced
phase space, for the variable $\varphi$ in this case.

Thus, we show that a solution of the WDW equation can coincide with the
wave function of the Dirac gaugeless quantization
in the reduced phase space where the wave function is normalizable and
describes the geometric evolution of the Universe for the Einstein comoving
frame of reference (\ref{psi}):
\be
\frac{d}{idT_F(a)}\left[\Psi^\pm_{WDW}(a,\varepsilon(n))e^{\mp iS(a)}\right]=
\frac{\varepsilon(n)}{a(T_F)}
\left[\Psi^\pm_{WDW}(a,\varepsilon(n))e^{\mp iS(a)}\right].
\ee
We see that variation of such a WDW function with respect to the Friedmann
time $T_F$ gives the "observable" Friedmann energy  (\ref{psi}).

\section{\p {Interpretation and conclusion.}}

The aim of the present paper is to investigate relations between the
Friedmann cosmological observables and the Dirac physical ones
in the Hamiltonian approach to quantization of the Universe using a simple
but important example of the homogeneous Universe filled in by the scalar
field excitations.

An essential difference of the research presented here from the analogous
papers on the Hamiltonian dynamics of cosmological models is complete
separation of the sector of physical invariant variables from the pure gauge
sector by the application of the gaugeless reduction ~\cite{PhRevD,JMPh}.
The main point is that in the process of the reduction
one of variables converts in to the observable invariant time.
We have shown that this conversion of the variable to the time parameter
leads to the normalizability of the Wheeler -- DeWitt (WDW) wave
function and removes an infinite factor in the Hartle -- Hawking
functional integral. The gaugeless reduction gives us the definite
mathematical and physical treatment of the WDW wave function and clears up
its relation to the observational cosmology.

The obtained wave function of the Friedmann Universe filled in by the
homogeneous scalar field is nothing but the one of
the scalar field excitations in the finite cavity of the conformal space
occupied by the Universe. The time evolution is governed by
the reduced Hamiltonian that coincides with the
conventional Hamiltonian for the massless excitation in field theory.

The considered gaugeless reduction distinguishes the conformal
(\cite{Narlik}) frame of reference. It was shown that, in both
the classical and quantum theories, the Friedmann observables
are connected with Dirac ones by the conformal
transformations with the cosmic  scale factor.

\vspace{0.5cm}
Acknowledgments.

We are happy to acknowledge interesting and critical
discussions with Profs. Cecile DeWitt-Morette, Z.Perjes
and to thank the Russian Foundation
for Basic Investigation, Grant N 96\--01\--01223 for support.


\begin{thebibliography}{99}
\bibitem{Dir58} P.A.M.~Dirac. Proc.Roy.Soc., \underline {A~246} (1958) 333;
Phys.Rev. \underline{114} (1959) 924.
\bibitem{ADM} R.~Arnowitt, S.~Deser, C.W.~Misner. Phys.Rev.
\underline {117} (1960) 1595.
\bibitem{Wheel} J.A.Wheeler. In Batelle Recontres :
1967 Lectures  in Mathematics and Physics, edited by  C. DeWitt and
J.A.Wheeler, Benjamin, New York, (1968).
\bibitem{DeWitt} B.S.~DeWitt. Phys.Rev. \underline{160} (1967) 1113.
\bibitem{Faddeev} L.D.~Faddeev,V.N.~Popov. Usp. Fiz. Nauk 111 (1973) 427.
\bibitem{Ryan1} M.P.~Ryan, Jr., and L.C.~Shapley. "Homogeneous Relativistic
Cosmologies", Princeton Series on Physics, Princeton University
Press, Princeton, N.Y. 1975.
\bibitem{Ryan2}M.P.~Ryan, "Hamiltonian Cosmology",
Lecture Notes in Physics N 13 Springer Verlag,
Berlin--Heidelberg--New York, 1972.
\bibitem{PhRevD} S.A. Gogilidze, A.M. Khvedelidze, V.N. Pervushin.
Phys. Rev.D 53 (1996) 2160.
\bibitem{JMPh} S.A. Gogilidze, A.M. Khvedelidze, V.N. Pervushin.
J.Math.Phys. 37 (1996) 1760.
\bibitem{natur1} J.P. Ostriker. P.J. Steinhard, Nature, Vol. {\bf 377},
19 Oct. 1995.
\bibitem{natur2}  R.C. Kennicutt Jr, Nature, Vol. {\bf 381},
13 June 1995.
\bibitem{Fried} A.A.~Friedmann, Z. Phys. 10 (1922) 377.
\bibitem{Narlik} J.V.~Narlikar in "Astrofizica e  Cosmologia, Gravitazione,
Quanti e Relativita", G. Barbera, Firenze , 1979.
\bibitem{Zeld} Ja.B. Zel'dovitch, Usp. Fiz. Nauk. 80 (1963) 357.
(in Russian).
\bibitem{Staniuk} K.P. Staniukovch, V.N. Melnikov, "Gidrodinamika, pola i
konstanty v teorii gravitatsii, Moscow, Energoatomizdat, 1973, p.105.
(in Russian)
\bibitem{Dirac} P.A.M.~Dirac, Lectures on Quantum Mechanics, Belfer Graduate
School of Science Yeshiva University, New York, 1964.
\bibitem{PhLet} V.Pervushin, V.Papoyan, S.Gogilidze, A.Khvedelidze, Yu.Palii,
V.Smirichinski, Phys.Lett.{\bf B365}(1996) 35
\bibitem{Towmas}
V.Pervushin, T.Towmasjan. Int.J.Mod.Phys.{\bf D4}(1995) N 1, 105-113.
\bibitem{Khved} A.M.Khvedelidze, V.V.Papoyan, V.N.Pervushin. Phys.Rev.D
{\p \bf 51}, (1995) 5654.
\bibitem{Rubakov} G.~Lavrelashvili, V.A.~Rubakov, P.G.~Tinyakov,
 Proceed. of the 5 seminar Quantum Gravity, Moscov, USSR, 28 May-1 June 1990,
edit. M.A.~ Markov et all.
\bibitem{hart}
J.B. Hartle, S.W. Hawking, Phys. Rev. D {\p \bf 28} (1983), 2960.
\bibitem{Hyward} S.A.~Hayward, Phys. Rev. D 53 (1996) 5664.

\end{thebibliography}
\end{document}